\documentclass{elsart}
\usepackage{amsmath,amssymb,epsfig,psfig}
\def\be{\begin{eqnarray}}
\def\ee{\end{eqnarray}}
\def\pvec{{\mbox{\boldmath $p$}}}

\def\bq{{\mbox{\boldmath $q$}}}

\def\qvec{{\mbox{\boldmath $q$}}}
\def\kvec{{\mbox{\boldmath $k$}}}

\def\sigmavec{{\mbox{\boldmath $\sigma$}}}

\newcommand{\anu}{\bar\nu}
\newcommand{\omnu}{\omega_{\bar\nu}}
\newcommand{\ome}{\omega_{e}}
\newcommand{\sla}{\! \not \!}
\newcommand{\ep}{\epsilon}
\newcommand{\Real}{\Re{\rm e~}}
\newcommand{\Img}{\Im{\rm m}}

\begin{document}
\begin{frontmatter}
\title{Direct Urca neutrino radiation from superfluid baryonic matter}
\author{Armen Sedrakian}
\address{ Institute for Theoretical Physics, T\"ubingen University,\\
D-72076 T\"ubingen, Germany}
\date{\today}
 
\begin{abstract}                           
Dense matter in compact stars cools efficiently by 
neutrino emission via the direct Urca processes $n\to p + e+\anu_e$ 
and $p+e\to n+\nu_e$. Below the pairing phase transition 
temperature $T_c$ these processes are  suppressed - 
at asymptotically low temperatures exponentially. We compute the 
emissivity of the Urca process at one loop for temperatures 
$T\le T_c$, in the case where the baryons are paired in 
the $^1S_0$ partial wave.
The Urca process is suppressed linearly, rather than exponentially, 
within the temperature range $0.1 \le T/T_c\le 1$. The charge-current Cooper 
pair-breaking process contributes up to $50\%$ of the 
total Urca emissivity  in this temperature range.
\end{abstract}
\end{frontmatter}
%\pacs{PACS 21.65.+f, 95.30.Cq; 97.60.Jd}

\section{Introduction and summary}

The direct Urca process (single neutrino $\beta$-decay) 
was introduced in astrophysics by Gamow and Schoenberg in their studies 
of stellar collapse and supernova explosions \cite{GAMOW}.
In a newly born neutron star,
the proton fraction drops to a few 
percent of the net density at the end of  the deleptonization phase.
At such low concentrations of charged particles the energy and 
momentum can not be conserved in a weak charged current decay of a 
single (on-shell) particle because of the large mismatch between 
the Fermi energies of baryons, therefore  
the direct Urca process is forbidden.
The two-baryon processes are not constrained 
kinematically~\cite{SALPETER}, and the so-called modified 
Urca process $n+n\to n+p+e+\anu_e$ is the main source of neutrino emission at 
low proton concentrations. Boguta \cite{BOGUTA} and 
Lattimer et al.~\cite{LATTIMER} pointed out that the high density 
matter can sustain large enough proton fraction  ($\ge 11\%-13\%$
according to ref.~\cite{LATTIMER}) for the Urca process to work. 
The proton fraction in dense matter is 
controlled by the symmetry energy of the  neutron rich nuclear matter which
is not well constrained. Exotic states of matter open new channels of 
rapid cooling with neutrino emission rates
comparable to the direct Urca rate \cite{PETHICK}.

Below $T_c$  pair correlations open a gap ($\Delta$) in the
quasiparticle spectrum of baryons\footnote{Pairing in 
higher than $l=0$ angular 
momentum states can lead to quasiparticle spectra featuring sections of 
Fermi surfaces where the gap vanishes; here
we shall treated only $l=0$, $S$-wave pairing in the 
$^1S_0$ partial wave. The $S$-wave pairing in the 
$^3S_1-^3D_1$ channel is suppressed by the mismatch in the
Fermi-energies of baryons required by the Urca process
~\cite{SEDRAKIAN_LOMBARDO} and can be ignored in the present 
context.}.  At low temperatures, when
the width of the energy states accessible to quasiparticles due to the 
temperature smearing of the Fermi-surfaces is less than the
gap in the quasiparticle spectrum, the neutrino production by the
direct Urca process is  
quenched by a factor ${\rm exp}(-\Delta/T)$, where $\Delta$ is the
larger of the neutron and proton gaps.
At moderate temperatures $T\le T_c$,  the changes in the 
phase-space occupation caused by the gaps in the quasiparticle spectra 
are non-exponential;  and the weak decay matrix elements 
acquire coherence pre-factors, since the degrees of freedom
(the Bogolyubov quasiparticles) are a superposition
of the particles and holes describing the unpaired state. The phase space 
changes due to the pairing were cast in suppression factors 
multiplying the emissivity of the processes  in the normal state 
in refs.~\cite{YAKOVLEV}; the role of coherence factors
has not been studied so far.

Our aim below is to compute the direct Urca emissivity in 
the superfluid phases of dense nucleonic matter below $T_c$.
We start by expressing the neutrino emissivity of the direct Urca process  
in terms  of the polarization tensor of matter within 
the real-time Green's functions approach to the 
quantum neutrino transport~\cite{SEDRAKIAN}. (The derivation in
Section 2 applies to a larger class of charge current reactions involving
$\beta$-decay; an example is the modified Urca process). 
The polarization tensor 
is then computed at one-loop for the isospin asymmetric nuclear matter
where the $S$-wave pairing is among the same-isospin 
quasiparticles. Within the range of temperatures $0.1 \le T/T_c\le 1$
the suppression of the Urca process will turn out to be linear 
rather than exponential in temperature.  An additional contribution to 
the Urca process comes from (previously ignored) charge-current 
pair-breaking process. Both  the charge-neutral current counterparts
of the  pair-breaking processes~\cite{FLOWERS,VOSKRESENSKY1,KAMINKER} 
and the direct Urca processes (suppressed by pair-correlations below $T_c$)
were found important in numerical  simulations of compact 
star cooling~\cite{SCHAAB1,SCHAAB2,PAGE1,PAGE2,PAGE3,KAMIN,TSURUTA,BLA}. 
Below,  we shall treat the baryonic matter as non-relativistic Fermi-liquid; 
one should keep in mind, however, that the relativistic mean-field treatments 
differ quantitatively from their 
non-relativistic counterparts~\cite{LEINSON1,LEINSON2}.

\section{Propagators and self-energies}

Stellar matter is approximately in equilibrium with respect to the weak 
processes - the $\beta$-decay and electron capture rates are nearly
equal. Thus, we need to compute the rate of, say,
anti-neutrino production by neutron $\beta$-decay and multiply the final
result by a factor of 2. We start with the transport equation 
for anti-neutrinos~\cite{SEDRAKIAN}
\be\label{BE_ANU}
\left[\partial_t + \vec \partial_q\,\omnu (\bq) \vec\partial_X
\right] f_{\anu}(\bq) =
\int^{0}_{-\infty} \frac{dq_0}{2\pi} {\rm Tr} 
\left[\sigma^<(q)S_{\anu}^>(q)-\sigma^>(q)S_{\anu}^<(q)\right],
\ee
where $f_{\anu}(\bq)$ is the anti-neutrino Wigner function,
$S_{\anu}^{>,<}(q)$ and $\sigma^{>,<}(q)$  are their propagators and
self-energies, $\omnu(\bq)=\vert \bq_{\anu}\vert$ is the on-mass-shell
anti-neutrino energy and $q = (q_0,\bq)$ is the four-momentum. 
The Wigner function, propagators and self-energies depend on the 
center-of-mass coordinate $X$, which is implicit in
Eq. (\ref{BE_ANU}).  If the leptons are on-shell and massless, 
their propagators are related to their Wigner functions via 
the quasiparticle ansatz \cite{SEDRAKIAN}
\be\label{PROP1}
S_{\lambda}^<(q)&=& \frac{i\pi\sla q}{\omega_{\lambda}(\bq)}
\Big\{\delta\left(q_0-\omega_{\lambda}(\bq)\right)f_{\lambda}(q)
-\delta\left(q_0+\omega_{\bar\lambda}(\bq)\right) 
\left[1-f_{\bar\lambda}(-q)\right] \Big\},
\ee
where the index $\lambda = \nu,\,\,e$ refers to neutrino ($\nu$) and
electron ($e$) and $\bar\lambda = \anu,\,\,\bar e$ to anti-neutrino 
($\anu$) and positron ($\bar e$). Both neutrinos and electrons 
obey a linear dispersion relation $\omega_{\lambda} = \vert\bq_{\lambda}\vert$,
since the electron neutrino mass can be ignored on 
energy scales $\sim 10^2$ keV, while 
the electrons are ultrarelativistic at relevant densities. In addition to 
the propagator $S_{\lambda}^<(q)$ we will need the ``time-reversed''
propagator $S_{\lambda}^>(q)$ which is obtained from (\ref{PROP1}) 
by an interchange of the particle and hole distributions 
$f_{\lambda}(q)\leftrightarrow 1-f_{\lambda}(q).$
\begin{figure}[tx]
\begin{center}
\mbox{\psfig{figure=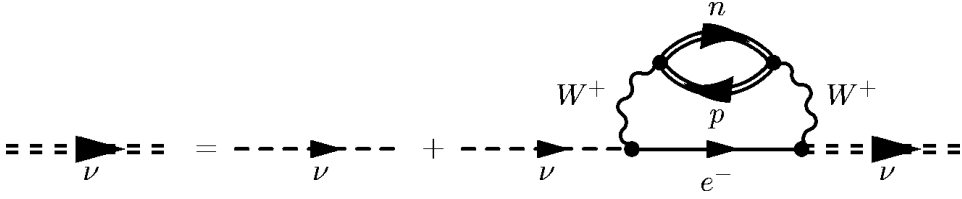,height=1.in,width=5.in,angle=0}}
\end{center}
\caption[]
{\small The diagrammatic Dyson equation for neutrinos which defines the 
self-energies $\sigma^{>,<}(q)$. The free and full neutrino propagators
are shown by the dashed and double dashed lines, the electrons and 
paired baryons  - by the solid and double-solid lines;
the wavy lines correspond to the gauge $W^{+}$-boson propagator. 
According to Eq. (\ref{PROP1}),  the neutrino Green's function 
simultaneously propagates a neutrino and anti-neutrino hole; the
time-reversed Green's function propagates a anti-neutrino and a
neutrino hole.
}
\label{fig1}
\end{figure}
The standard model neutrino-baryon charged-current interaction is 
\be\label{HAM}
V = \frac{\tilde G}{\sqrt{2}} J^{(H)} \, J^{(L)}, \quad
J^{(H)}_{\mu} =\overline \phi \gamma_{\mu}
(c_V - c_A\gamma_5) \phi ,\quad J_{(L)}^{\mu}
= \overline\psi \gamma^{\mu}(1-\gamma_5)\psi ,
\ee
where $\tilde G \equiv G_F \cos \theta$ with $G_F$ being the weak 
coupling constant and $\theta$ the Cabibbo angle ($\cos \theta = 0.973$),
$\psi$ and $\phi$ are the neutrino and baryon field operators, 
$c_V$ and $c_A$ are the weak charged-current vector and 
axial vector coupling constants.
The neutrino self-energies $\sigma^{>,<}(q)$ can be evaluated
perturbatively with respect to the weak interaction. 
To lowest  order in $\tilde G$ the second-order Born diagram 
gives (see Fig. 1)
\be
-i\sigma^{>,<}(q_1) &=&
\sum_{q,q_2} \delta^4(q_1 - q_2 + q)~ i\Gamma_{(L)}^{\mu}~
iS_e^{<}(q_2)~ i\Gamma_{(L)}^{\dagger\, \zeta}
~i \Pi^{<,>}_{\mu\zeta}(q),
\ee
where $\sum_q\equiv \int d^4q/(2\pi)^4$ is the phase-space 
integration symbol, $\Pi^{>,<}_{\mu\zeta}(q)$ is the 
baryon polarization tensor and $\Gamma_{\mu (L)} =
\gamma_{\mu}(1-\gamma_5)$  is the weak interaction vertex, 
which is independent of $q$ at the energy and 
momentum transfers much smaller than the gauge boson mass. 
Substituting the self-energies and the propagators in the collision
integral we find for the loss part [the second term on the r.-h. side
of Eq. (\ref{BE_ANU})]
\be\label{LOSS}
&&I_{\anu}^{>}(\bq_1)= i\int^{0}_{-\infty}\!\! \frac{dq_{10}}{2\pi}  
\sum_{q,q_2}\delta^4(q_1 - q_2 + q)~{\rm Tr}\left[~\Gamma^{\mu}_{(L)}
\frac{\pi\sla q_2}{\omega_e(\bq_2)}\Gamma^{\dagger\,\zeta}_{(L)}
\frac{\pi\sla q_1}{\omnu(\bq_1)}\right]\nonumber\\
&&\delta\left(q_{02}-\omega_e(\bq_2)\right)\left[1-f_{e}(q_2)\right]
\delta\left(q_{10}+\omnu(\bq_1)\right)\left[1-f_{\anu}(q_1)\right]
\Pi_{\mu\zeta}^{<}(q),
\ee
where we retained from the collision integral the only relevant 
term which has an electron and anti-neutrino in the final state.
The gain term naturally does not contribute to the emission rate.

\section{Direct Urca emissivities}

The anti-neutrino emissivity (the energy radiated per 
unit time) is obtained from Eq. (\ref{BE_ANU}) 
\be
\epsilon_{\anu}&=&\frac{d}{dt}\int\!\frac{d^3q}{(2\pi)^3}
f_{\anu}(\bq)\omnu(\bq).
\ee
Carrying out the energy integrations in Eq. (\ref{LOSS}) 
and using the equilibrium identity  
$\Pi_{\mu\zeta}^{<}(q) = 2i g_B(q_0) {\Img}\,
\Pi_{\mu\zeta}^R(q)$, where $g_B(q_0)$ is the Bose distribution 
function and $\Pi^R_{\mu\zeta}(q)$ is the retarded polarization 
function, we obtain
\be\label{EMISSIVITY}
\epsilon_{\anu}&=& - 2\left( \frac{\tilde G}{\sqrt{2}}\right)^2
\int\!\frac{d^3q_1}{(2\pi)^32 \ome(\qvec_1)}
\int\!\frac{d^3 q_2}{(2\pi)^3 2\omnu(\qvec_2)}
\int\! d^4 q \,\delta (\qvec_1 + \qvec_2 -  \qvec)
\nonumber\\
&&\delta(\ome+\omnu-q_{0})\, \omnu(\qvec_2)
 g_B(q_0)\left[1-f_{e}(\ome)\right]
 \Lambda^{\mu\zeta}(q_1,q_2)\Img\,\Pi^R_{\mu\zeta}(q),
\ee
where $\Lambda^{\mu\zeta}(q_1,q_2) = {\rm Tr}\left[\gamma^{\mu}
(1 - \gamma^5)\sla q_1\gamma^{\zeta}(1-\gamma^5)\sla q_2\right]$.
The symbol $\Img$ refers to the imaginary part the polarization 
tensor's resolvent.
Since the baryonic component of stellar matter is in thermal equilibrium 
to a good approximation, it is convenient to use the Matsubara Green's 
functions (ref. \cite{ABRIKOSOV}, pg. 120)
\be
G_{\sigma\sigma'}(\pvec ,\tau) &=& -\delta_{\sigma\sigma'}
\langle T_{\tau} a_{p\sigma}(\tau)a^{\dagger}_{p\sigma'}(0)\rangle ,
\nonumber \\
F_{\sigma\sigma'}(\pvec ,\tau) &=& 
\langle T_{\tau} a_{-p\downarrow}(\tau)
a_{p\uparrow}(0)\rangle ,\quad \quad
F^{\dagger}_{\sigma\sigma'}(\pvec ,\tau) 
=\langle T_{\tau} a_{p\uparrow}^{\dagger}(\tau)
a^{\dagger}_{-p\downarrow}(0)\rangle ,
\ee
where $\tau$ is the imaginary time, $a^{\dagger}_{p\sigma}(\tau)$ and
$a_{p\sigma}(\tau)$ are the creation and destruction operators, 
$\sigma = \uparrow , \downarrow$ stands for spin and the propagators 
are diagonal in the isospin space.
The quasiparticle spectra of Cooper pairs coupled in a relative 
$S$-wave state are (the wave-vectors $p$ and $k$ refer
to protons and neutrons)
$\ep_p = \sqrt{\xi_p^{2}+\Delta_p^{2}}$,
$\ep_k = \sqrt{\xi_k^{2}+\Delta_n^{2}}$
with (omitting the isospin index) $\xi = {p^2}/{2m}+{\Real}\Sigma (p)
-\mu$, where $\mu$, $m$ and $\Sigma(p)$ are the chemical 
potentials, masses and self-energies of protons and neutrons. 
Expanding the self-energy around the Fermi-momentum $p_F$ leads to
$ \xi = {p^2}/{2m^*} - \mu^*, $ where 
$\mu^* = -{\Real}\Sigma(p_F)+\mu, $ and 
$m/m^* = (1+\partial \Real\Sigma(p)/\partial p \vert_{p=p_F})$.
\begin{figure}
\begin{center}
\mbox{\psfig{figure=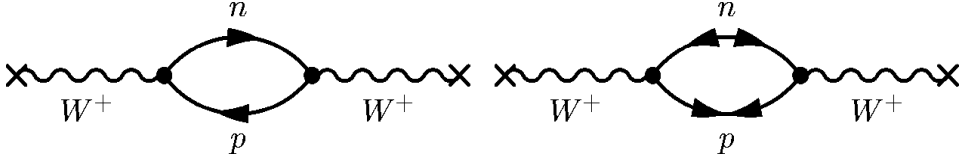,height=.8in,width=5in,angle=0}}
\end{center}
\caption[]
{\footnotesize{
The one loop baryon polarization tensor. The baryon and (amputated) 
$W^+$-boson propagators are shown by the solid and wave lines.
The double arrow lines refer to the anomalous propagators 
(ref. \cite{ABRIKOSOV}, pg. 300}).
}
\label{fig2}
\end{figure}
The density and spin-density response functions are defined 
in terms of imaginary-time ordered products 
($T_{\tau}$ is the ordering symbol)
\be
\Pi (\qvec, iq_0) &=&\sum_{\sigma,\sigma',p, p'}
\int_0^{\beta}d\tau e^{i\omega\tau} \alpha\cdot \alpha'
\langle T_{\tau} a^{\dagger}_{p+q,\sigma}(\tau)a_{p,\sigma}(\tau)
a^{\dagger}_{p'-q,\sigma'}(0)a_{p',\sigma'}(0)\rangle ,
\ee
where $\alpha =1 $ for the  vector and $\sigmavec$ for the axial vector 
response ($\sigmavec$ is the vector of  Pauli matrices).
The one-loop density-density and spin-spin correlation functions 
are given by 
\be\label{PiV1} 
\Pi_{V/A} (\qvec, iq_0) =\frac{1}{\beta}\sum_{\sigma p} 
\left[G(\pvec,ip)G(\kvec,ik_0)
\mp F(\pvec,ip)F^{\dagger}(\kvec, ik_0) \right],
\ee
where $p = (ip_0, \pvec)$,  $k = (ip_0+iq_0, \pvec+\qvec)$, 
the zeroth components of these four-vectors are the complex
fermionic Matsubara frequencies, and  $\beta$ is the inverse
temperature. The different signs in Eq. (\ref{PiV1}) reflect the 
fact that under time reversal the $\alpha$ vertex is even 
for scalar and odd for spinor perturbations. Note that the one-loop
approximation above ignores the vertex corrections, which can 
be treated within the Fermi-liquid theory~\cite{VOSKRESENSKY2}.
The summation over the Matsubara frequencies in (\ref{PiV1}) followed
by analytical continuation gives
\be\label{eq:4}
\Pi_{V/A}^R (\qvec, \omega)&=&
\sum_{\sigma, \vec p}\Biggl\{
\left(\frac{u_p^2u_k^2\mp u_pu_kv_pv_k}{\omega+\ep_p-\ep_k+i\delta}
-\frac{v_p^2v_k^2\mp u_pu_kv_pv_k}{\omega-\ep_p+\ep_k+i\delta}
\right)\left[f(\ep_p)-f(\ep_k)\right]\nonumber\\
&&\hspace{-1.5cm}+
\left(\frac{u_p^2v_k^2\pm u_pu_kv_pv_k}{\omega-\ep_p-\ep_k+i\delta}
-\frac{u_p^2v_k^2\pm u_pu_kv_pv_k}{\omega+\ep_p+\ep_k+i\delta}
\right)\left[f(-\ep_p)-f(\ep_k)\right]\Biggr\},
\ee
where 
$u_p^2 = (1/2)\left(1+\vert \xi_p\vert/\ep_p\right)$ and $u_p^2+v_p^2 = 1$.
The first term in Eq. (\ref{eq:4}) describes the quasiparticle 
scattering, which is the generalization to superfluids 
of the ordinary scattering in the unpaired state; the second one describes 
pair-breaking, which is specific to superfluids.
Taking the imaginary part  and changing to dimensionless variables
$y=\beta\omega$, $x = \beta \xi_p$ and  $w_i = \beta \Delta_i$
($i = p,n$)
we obtain
\be\label{POLAR} 
{\rm Im}\Pi_{V/A}^R(q,y)&=& -\frac{m_n^*m_p^*}{2\pi q\beta} 
\left[{I}^{SC}_{V/A}(y)+{I}^{PB}_{V/A}(y)\right],
\ee
where the (dimensionless) integrals for the vector (V) and axial vector
(A) couplings are
\be
\label{Iintergrals}
{I}^{SC/PB}_V(y) &=&\int_{-\infty}^{\infty}dx~
{C}^{SC/PB}_{V}(x,y,w_p)\vert \Xi^{SC/PB}(x,y,w_n)\vert\nonumber\\
&\times& \left[f(\pm\sqrt{x^2+w_p^2})-f(\sqrt{x^2+w_p^2}+y)\right]
\theta (1-\vert x_0^{\pm}\vert) .
\ee
where $\Xi^{SC/PB}(x,y,w_n) = (\omega\pm\epsilon_p)\left\{\vert
  (\omega\pm\epsilon_p)^2-\Delta_n^2\vert\right\}^{-1/2}$, and
\be
x_0^{\pm}  = {\rm sgn}(\omega\pm\epsilon_p)
\frac{m_n^*}{pq}\left\{\sqrt{\vert (\omega\pm \ep_p)^2
-\Delta_n^2\vert }-l^2\right\}
\ee
with $l^2\equiv (p^2+q^2)/(2m_n^*)-\mu_n^*$.
The axial-vector integrals  ${I}^{SC/PB}_A$ are obtained
by replacing in ({\ref{Iintergrals}) 
the  coherence factors ${C}^{SC/PB}_{V}$ by ${C}^{SC/PB}_{A}$,
where $C^{SC}_{V/A} = (u_pu_k\mp v_pv_k)^2,$ and 
$C^{PB}_{V/A} =u_k^2v_p^2\pm u_pu_kv_pv_k.$ On substituting 
Eq.~(\ref{POLAR}) in Eq.~(\ref{EMISSIVITY}) we obtain
\be\label{SUP_EMISSIVITY}
\epsilon_{\anu}&=& - 8 \tilde G^2 
\frac{m_n^*m_p^*}{2\pi \beta}\int\! \frac{d^4 q }{q}
~g_B(\omega)\left[({  I}^{SC}_V+ {  I}^{PB}_V)
+ 3 g_A^2 ({  I}^{SC}_A+ {  I}^{PB}_A) \right]
\int\!\frac{d^3q_1}{(2\pi)^32 \ome}\nonumber\\
&\times&\int\!\frac{d^3 q_2}{(2\pi)^3 2\omnu }
\left[1-f_{e}(\ome)\right]\delta^3(\qvec_1 + \qvec_2 -  \qvec)
\delta(\ome+\omnu-\omega)\, \omnu^2\ome ,
\ee
where the (unquenched) axial coupling constant $g_A= 1.26$. Since 
the neutrino momentum is much smaller than the electron Fermi-momentum,
$\qvec \simeq \qvec_1$ and $\vert q_1\vert\simeq p_{Fe}$.
With this approximation our final result is
\be\label{SUP_EMISSIVITY4}
\epsilon_{\anu}&=&  
\frac{3 \tilde G^2 m_n^*m_p^* p_{Fe}}{2\pi^5 \beta^6}J= \epsilon_0 J,\\
J &=&
-\frac{1}{6}\int_{-\infty}^{\infty}\!dy~g_B(y)
\left[({  I}^{SC}_V+ {  I}^{PB}_V)
+ 3 g_A^2 ({  I}^{SC}_A+ {  I}^{PB}_A) \right]\int_{0}^{\infty} dz z^3
f_e(z-y).\nonumber
\ee
The second integral can be expressed through the polylogarithmic
function ${\rm Li}_4[-{\rm exp}(y)]$, though this is not particularly 
illuminating. Above $T_c$ Eq. (\ref{SUP_EMISSIVITY4}) 
yields the finite-temperature emissivity of the direct Urca process
\be\label{URCA_FINITE_T}
\epsilon_{\anu}=  (1+3g_A^2)\epsilon_0
\int\!dy~g_B(y)~{\rm ln} \frac{1+e^{-x_{\rm min}}}
{1+e^{-(x_{\rm min}+y)}}\int dz z^3f_e(z-y) ,
\ee
where  $x_{\rm min} = \beta (p_{\rm min}^2/2m^* -\mu_p^*)$ and 
$p_{\rm min} = (m^* /q)(\omega-\mu_p^* +\mu_n^* -q^2/2m^*)$ 
[here we assume $m_n^* \simeq m_p^*   = m^* $]. At zero temperature 
the logarithm  in Eq. (\ref{URCA_FINITE_T}) reduces to 
$y\theta(-x_{\rm min})$, the integrals are analytical and one 
easily recovers the zero-temperature result of ref. \cite{LATTIMER}.
\begin{figure}[t]
\begin{center}
\mbox{\psfig{figure=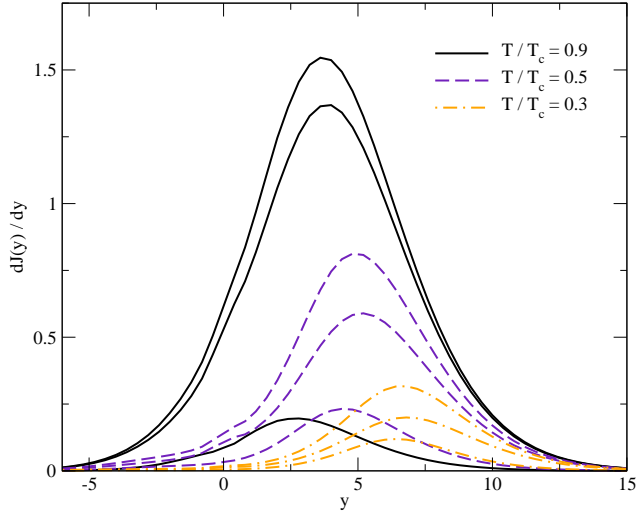,height=4.5in,width=3.5in,angle=-90}}
\end{center}
\caption[]
{\small{The energy distribution of the 
neutrino radiation $dJ(y)/dy$  for $T/T_c = 0.9$ (solid lines), 0.5 (dashed 
lines), 0.3 (dashed dotted lines). For each fixed $T/T_c$ the lines
correspond to (bottom to top) the pair-breaking, scattering and 
total contributions.
}}
\label{fig4}
\end{figure}
\begin{figure}[htb]
\begin{center}
\mbox{\psfig{figure=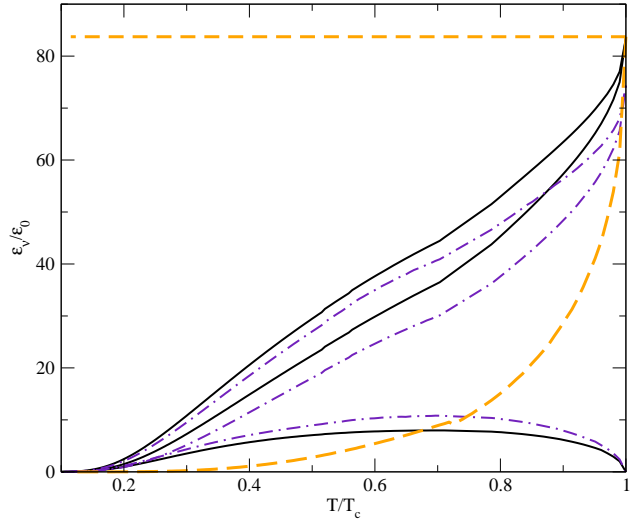,height=4.5in,width=3.5in,angle=-90}}
\end{center}
\caption[]
{\footnotesize{The neutrino emissivity in units 
of $\epsilon_0$ versus temperature (solid lines $\Delta_n(0)=\Delta_p(0) =
0.5$ MeV, dashed-dotted lines $\Delta_n(0)=0.5,$ $\Delta_p(0) = 2$ MeV). 
The scattering, pair-breaking
contributions and their sum are shown by  dashed and 
dashed-dotted and solid lines. The upper short-dashed line is the 
extrapolation of the rate for unpaired matter to low temperatures, the lower 
one corresponds to the exponential suppression as discussed in the text.
}}
\label{fig4}
\end{figure}
\begin{figure}[t]
\begin{center}
\mbox{\psfig{figure=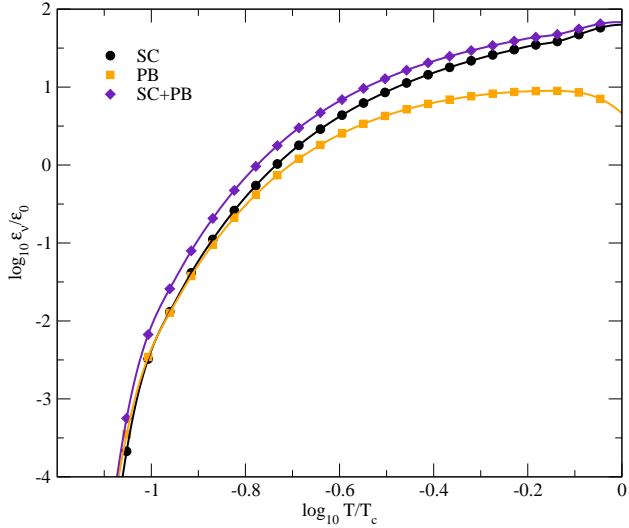,height=4.5in,width=3.5in,angle=-90}}
\end{center}
\caption[]
{\footnotesize{The log-log plot of the 
dependence of the neutrino emissivity in units 
of $\epsilon_0$ on temperature. The scattering, pair-breaking
contributions and their sum are marked  by circles, squares, and
diamonds respectively.
}}
\label{fig5}
\end{figure}

\section{A model calculation}
To set-up an illustrative model,  we assume
$\Delta_n(T=0) = \Delta_p(T=0) = 0.5$~MeV, $m_n^*/m_n = m_p^*/m_p = 0.7$,
$\Real\Sigma (p_{Fn}) = -10$~MeV and $\Real\Sigma (p_{Fp}) = -160$~MeV
at the density 0.24 fm$^{-3}$, which is above the Urca threshold in 
charge neutral nucleonic matter under $\beta$-equilibrium for the 
assumed values of the baryon self-energies.
The neutron self-energy shift is taken more repulsive 
than the one found in non-relativistic calculations with 
two-body interactions to mimic the effect of repulsive 
three-body forces. The remaining parameters of the model are 
$p_{Fn} =    1.8$ fm$^{-1}$,  $p_{Fp} = p_{Fe}=   1.07$ fm$^{-1}$, 
$\mu_n^* = 96.3$ MeV, $\mu_p^* = 34.2$ MeV  
and  $\mu_e = 212.1$ MeV.

Fig.~3 shows the energy distribution of the neutrino radiation
$dJ(y)/dy$ for various temperatures. The energy distribution is 
thermal with the maximum at $\omega\sim 3T$ in agreement with 
the fact that each anti-neutrino carries on average 
energy  equal to $3T$ in units of Boltzmann constant. For smaller $T/T_c$
the maximum shifts to larger energies corresponding to higher 
``effective temperature'' and the peak of the distribution is 
reduced because of pairing; the thermal shape of the distribution 
remains unchanged.
While in the limit $T/T_c\le 1$ the scattering  contribution
dominates,  at lower temperatures the
pair-breaking process becomes increasingly important; e.g. 
at $T/T_c \sim 0.3$ it contributes about the half of the scattering 
contribution.  

Fig.~4 focuses on the temperature dependence of the direct 
neutrino emissivity in the range  $0.1\le T/T_c\le 1$.
The important feature here is the nearly 
linear dependence of the emissivity on the temperature in the range
$0.1\le T/T_c\le 1$; the commonly assumed exponential decay  
- a factor exp($-\Delta/T$) with $\Delta = {\rm max}(\Delta_n,
\Delta_p)$  -  underestimates the emissivity. 
(Similar conclusion concerning the suppression of the 
direct Urca process by pair correlations was reached in
ref.~\cite{YAKOVLEV} which treated the scattering contribution to 
the emissivity).
The contribution of the pair-breaking processes becomes substantial
in the low-temperature range $0.1\le T/T_c\le 0.4$ (about the half of 
the scattering contribution at  $T/T_c\sim 0.3$). Increasing the 
value of the proton gap to 2 MeV while keeping the neutron gap at 
its value 0.5 MeV suppresses the emissivity of the scattering
process, since at a given temperature the phase space accessible to
the excited states is reduced. The pair-breaking processes are almost 
unaffected since they are related to the scattering of particles in
and out-of the condensate. The temperature 
dependence of the net emissivity can be crudely approximated as 
$\epsilon_{\anu} = 90\,\epsilon_0\, (T/T_c - 0.1)$; this reduces
to the standard Urca result for $T=T_c$ and sufficiently well (for 
the purpose of cooling simulations) reproduces the numerical 
result in Fig. 3 for $T/T_c\in[0.1;1]$.  Alternatively the rate of 
the Urca process in the unpaired matter can be suppressed by a 
factor $(10/9) (T/T_c - 0.1)$.
Fig~5 illustrates the low-temperature asymptotics of the emissivities
for  $\Delta_n=\Delta_p = 0.5$ MeV. For $T/T_c \le 0.1$ the
contribution of the pair-breaking process to the total emissivity 
is equal or larger than that of the scattering
one. Asymptotically, the logarithmic derivative of the emissivity, 
$d{\rm log}\, \epsilon_{\nu} /d{\rm log}\, T$, tends to a constant
(negative) value which depends only on the magnitude(s) of the zero 
temperature gap(s), as should be the case for exponentially 
suppressed emissivities.

\section*{Acknowledgments} I am grateful to Thomas Dahm, 
Alex Dieperink, Herbert M\"uther, 
Dany Page, Chris Pethick and Dimitry Voskresensky  for helpful 
conversations and correspondence and to the Institute for  
Nuclear Theory (Seattle)  and ECT$^*$ (Trento) for hospitality. 
This work was supported in part by the Sonder\-forschungs\-bereich 
382 of the Deutsche For\-schungs\-gemein\-schaft.


\begin{thebibliography}{99}
\bibitem{GAMOW} G. Gamow, M. Schoenberg,  Phys. Rev.  59 (1941) 539.
\bibitem{SALPETER} The modified Urca process was originally studied 
in  H. Y. Chiu, E. E. Salpeter, Phys. Rev. Lett. 12 (1964) 413;
J. N. Bahcall and R. A. Wolf, Phys. Rev. Lett. 14,  (1965) 343;
Phys. Rev. 140, B1452.
The recent work on these processes is described in 
D. N. Voskresensky, Lecture Notes in Physics vol. 578, Springer, Berlin, 2001,
p.~467 [{\tt  astro-ph/0009093}] and  D. G. Yakovlev, C. J. Pethick,
Ann. Rev. Astron. Astrophys. 42 (2004) 169.
\bibitem{BOGUTA}   J. Boguta, Phys. Lett. B 106 (1981) 255.
\bibitem{LATTIMER} J. M. Lattimer, C. J. Pethick, M. Prakash, P. Haensel,
                   Phys. Rev. Lett.  66 (1991) 2701.
\bibitem{PETHICK} C. J. Pethick, Rev. Mod. Phys. 64 (1992) 1133 and
  referencies therein.
\bibitem{SEDRAKIAN_LOMBARDO} A. Sedrakian and U. Lombardo, Phys. Rev. Lett.
                    84 (2000) 602.
\bibitem{YAKOVLEV}  K. P. Levenfish, D. G. Yakovlev,
                    Astron. Lett. 20 (1994) 43
                   [Pis'ma Astron. Zh. 20 (1994) 54]; D. G. Yakovlev,
K. P. Levenfish, Yu. A. Shibanov, Sov. Phys. Usp. 169 (1999) 825, Sec. 4.
\bibitem{SEDRAKIAN} A. Sedrakian and A. E. L. Dieperink, Phys. Rev. D 
                    62 (2000) 083002.
\bibitem{FLOWERS} E. G. Flowers, M. Ruderman, P. G. Sutherland,
  Astrophys. J. 205 (1976)  541.
\bibitem{VOSKRESENSKY1}D. N. Voskeresensky, A. V.  Senatorov,
 Sov. J. Nucl. Phys., 45 (1987) 411 [ Yad. Fiz. 45 (1987)  657].
\bibitem{KAMINKER} D. G. Yakovlev, A. D.  Kaminker, K. P.  Levenfish, 
                   Astron. Astrophys. 343 (1999) 650.
\bibitem{SCHAAB1} C. Schaab, D. Voskresensky, A. Sedrakian, F. Weber,
  M. Weigel, Astron.  Astrophys. 321 (1997) 591.
\bibitem{SCHAAB2}
C. Schaab,  A. Sedrakian, F. Weber,  M. Weigel, Astron.  Astrophys. 346 (1999) 465.
\bibitem{PAGE1} D. Page, In: {\it Neutron Stars and Pulsars},
  ed. N. Shibazaki, N. Kawai, S. Shibata, T. Kifune, p.183 (Universal
  Academy press, Tokyo 1998) [{\tt astro-ph/9802171}].
\bibitem{PAGE2} D.~Page, J.~M.~Lattimer, M.~Prakash, A.~Steiner, 
               Ap. J. suppl., in press, [{\tt astro-ph/0403657}].
\bibitem{PAGE3}  D.~Page, preprint   (2004)  [{\tt astro-ph/0405196}].
\bibitem{KAMIN} A. D. Kaminker, P. Haensel, D. G. Yakovlev,
  Astron. Astrophys. 383 (2001) L17.
\bibitem{TSURUTA}  S. Tsuruta, M. A. Teter, T. Takatsuka, T. Tatsumi, 
                    R. Tamagaki,  Ap. J. 571 (2002) L143.
\bibitem{BLA} D. Blaschke, H. Grigorian, D. Voskresensky, Astron. 
              Astrophys. 424 (2004) 979.
\bibitem{LEINSON1}   L. B. Leinson, A. P\'erez, Phys. Lett. B 518 (2001)
15; Phys. Lett. B 522 (2001) 358, Erratum.
\bibitem{LEINSON2} L. B. Leinson, Nucl. Phys. A 707 (2002) 543.
\bibitem{VOSKRESENSKY2}  D. N. Voskeresensky, A. V.  Senatorov, JETP 63 (1986) 
                       885 [Zh. Eksp. Teor. Fiz. 90 (1986) 1505].
\bibitem{ABRIKOSOV} A. A. Abrikosov, L. P. Gorkov, I. E. Dzyaloshinkski,
{\it Methods of quantum field theory in statistical physics}, (Dover, New
  York, 1975).
\end{thebibliography}
\end{document}